\documentclass[pra,twocolumn,a4paper,showpacs,superscriptaddress]{revtex4}

 \usepackage{latexsym}
 \usepackage{amsmath}
 \usepackage{amsfonts}
 \usepackage{graphicx}
 \usepackage{amssymb}
 \usepackage{subfigure}
 \usepackage{color}
 \usepackage{setspace}
 
 \newcommand{\ket}[1]{\ensuremath{|#1\rangle}}
 \newcommand{\bra}[1]{\ensuremath{\langle #1 |}}

 \newcommand{\bc}{\begin{center}}
 \newcommand{\ec}{\end{center}}
 \newcommand{\mf}[1]{\boldsymbol{#1}}
 
 \newcommand{\ii}{i}
 \newcommand{\DP}{\Delta_p}
 \newcommand{\DC}{\Delta_c}
 \newcommand{\DM}{\Delta_s}
 \newcommand{\wP}{\omega_p}
 \newcommand{\wC}{\omega_c}
 \newcommand{\wM}{\omega_s}
 \newcommand{\WP}{g}
 \newcommand{\WC}{G_{c}}
 \newcommand{\WM}{G_{s}}

\begin{document}

\title{ Kerr field induced tunable optical atomic  waveguide}

\author{Sandeep \surname{Sharma}}
\email{sandeep.sharma@iitg.ernet.in}
\affiliation{Department of Physics, Indian Institute of Technology Guwahati,
Guwahati- 781039, Assam, India}

\author{Tarak N. \surname{Dey}}
\email{tarak.dey@iitg.ernet.in}
\affiliation{Department of Physics, Indian Institute of Technology Guwahati,
Guwahati- 781039, Assam, India}
\date{\today}

\begin{abstract}
We propose an efficient scheme for the generation of tunable optical waveguide based on atomic vapor in $N$-type configuration. 
We exploit both control field induced transparency and Kerr field induced absorption to produce a flexible probe transparency window otherwise not feasible.
We employ a suitable spatial profile of control and Kerr beams to create a high contrast refractive index modulation that holds the key to guiding a weak narrow probe beam.
Further we numerically demonstrate that high contrast tunable waveguide permits the propagation of different modes of probe beam to several Rayleigh lengths without diffraction.
This efficient guiding of narrow optical beam may have important applications in large density image processing, and high resolution imaging.
\end{abstract}
\maketitle
\section{Introduction}
Controlling diffraction is a fundamental goal of image processing, high resolution imaging and optical lithography. 
Almost all finite width beams except Airy \cite{Berry79,Siviloglou07}, Bessel \cite{Durnin87,McGloina05}, Mathieu \cite{Vega00,Zhang12}, and parabolic beams \cite{Bandres} are subjected to diffraction while they propagate through free space or media. 
The diffraction induced distortion causes loss of information carried by the beam. 
Hence diffraction acts as a major obstacle in the generation, transfer, and processing of information.
In order to subdue the effect of diffraction, variety of techniques based on Electromagnetically Induced Transparency(EIT) \cite{Moseley,Truscott,Kapoor,Yavuz,Gorshkov,Li}, Coherent Population Trapping(CPT) \cite{Vudyasetu,Kiffner,Kapale,Dey}, and Saturated Absorption \cite{Dey1} have been proposed. 
In most of these schemes, the salient feature is to manipulate the susceptibility of the medium along the transverse direction using a spatially dependent control field. 
A suitable spatially dependent control field can create an optical waveguide like structure inside the atomic medium. 
The generated optical waveguide protect the probe field from diffraction induced distortion during propagation through the medium.
Moreover, the idea of manipulating the susceptibility of the medium along the transverse direction has been used for cloning of images, focusing, and self-imaging \cite{Moseley1,Cheng,Zhang,Verma,Ding,Cao}. 
However in all the above cases the minimum feature size of the guided beam is $\sim 50 \mu$m, which has a limitation for practical use in high resolution imaging, lithography, and image processing, due to its incapacity to process large density of information. 
Narrow feature size optical beams are essential to satisfy the above criterion. 
A high contrast waveguide is required to suppress or even eliminate diffraction of the narrow beam in the course of propagation. 
This waveguide may be realized by considering a medium with high optical density.
Simultaneously, the high optical density also increases the gain or absorption of the medium, which limits it practical applicability.

In order to overcome these limitations, we look for a system where the refractive index can be enhanced without a significant change of absorption or gain. 
The main motivation for the present study derives from recent  elegant experiments on photolithography by Scott et al. \cite{Scott} and Andrew et al. \cite{Andrew}. 
In this experiments, they used photochromic molecules as a mask to imprint a nanoscale pattern on a target material. 
The mask shows very interesting behaviour while interacting with light. 
For a light with wavelength $\lambda_{1}$ (writing beam) the mask becomes transparent whereas for another light with wavelength $\lambda_{2}$ (inhibitor beam) it becomes opaque. 
Exploiting these properties of the mask, they created a nano-size pattern using Gaussian beam (writing-beam) and a Laguerre-Gaussian (LG) beam (inhibitor-beam).  
Such interesting ideas have also been used in STED microscopy to image nano size particles \cite{Hell, Hell1}.  
A $N$-type four level system \cite{Dey_07,Tian} can be used as an atomic analog for the mask. 
The atomic system becomes transparent to the weak probe due to presence of a strong control field.
However it becomes opaque in the presence of an additional Kerr field.
The competition between control field induced EIT and Kerr field induced absorption along the transverse directions of the medium forms  a narrow atomic waveguide.
Also the Kerr field increases the variation of the refractive index between core and cladding which allows for distortionless propagation of narrow probe beam through several Rayleigh lengths.

In this paper, we demonstrate how a high contrast tunable optical waveguide can be facilitated by using additional Kerr field. 
To achieve this, we use a four-level $N$-type atomic system driven by probe, control and Kerr fields. 
We first derive an analytical expression for the probe susceptibility in the steady state limit. 
We find that the control field induced probe transparency can be abolished by a Kerr field at probe resonance. 
This Kerr field assisted probe absorption plays an important role in the generation of tunable waveguide. 
Next, we investigate the spatial inhomogeneity of the medium caused by transverse variation of both control and Kerr fields.  
We find that a Gaussian control beam creates a wide waveguide structure with low refractive index variation between core and cladding, inside the atomic medium. 
The waveguide structure is sensitive to the control beam size, however, even a narrow control beam cannot create a narrow waveguide due to its inherent diffraction.  
Inevitably this wide waveguide fails to protect the propagation of the narrow probe beam without diffraction. 
To avoid this limitation, we consider a LG Kerr field to facilitate a high contrast narrow waveguide structure.  
In order to delineate the effect of the high contrast waveguide on probe beam dynamics, we numerically solve the Maxwell's wave equation at the paraxial limit. 
Finally, we show that Kerr field induced tunable optical atomic waveguide can guide different modes of the probe beam to several orders of Rayleigh length without any diffraction.
This efficient guiding of narrow optical beam may have important applications in large density image processing and high resolution imaging.

This article is structured as follows.
We start  introducing the $N$-type four level system in Sec. II.A. 
We then derive the time independent Hamiltonian.
Next, we use the density matrix formalism to obtain the time dependent dynamical equation for evolution of atomic populations and coherences. 
In Sec. II.B, we adopt the perturbative approach to obtain an analytical expression for the medium susceptibility in the steady state limit. 
In Sec. II.C we describe the beam propagation equation under paraxial approximation. 
Successively, we use the Angular Spectrum Representation Method(ASRM) to confirm the validity of paraxial approximation.
In Sec. III.A, we study the effect of continuous wave (cw) control and Kerr field on medium susceptibility.
In Sec. III.B, we discuss the creation of a tunable waveguide by using spatially dependent control and Kerr fields. 
Finally, we provide the results from the beam-propagation equation in the presence of  high contrast tunable waveguide. 
In Sec. IV, we summarize our work.
\section{Theoretical Formulations}
\subsection{Model}
\begin{figure}[t!]
\includegraphics[width=8.5cm, height=5.6cm]{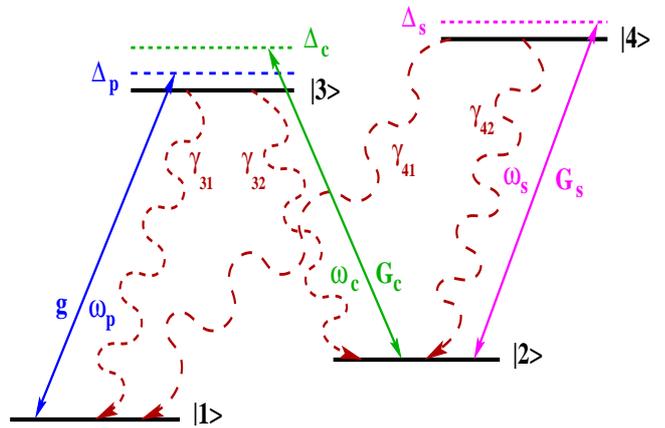}
\caption{\label{fig:Fig1} (Color online) 
Shows a schematic diagram of the level system. 
A probe field with Rabi frequency $\WP$ and a strong control with Rabi frequency $\WC$ couples the transitions $\ket{3}\leftrightarrow\ket{1}$ and $\ket{3}\leftrightarrow\ket{2}$, respectively. 
The transition $\ket{4}\leftrightarrow\ket{2}$ is coupled by a strong Kerr field having Rabi frequency $\WM$. 
$\Gamma_{3i}$ and $\Gamma_{4i}$, $i\in \{1,2\}$  are the radiative decay rates from the excited states $\ket{3}$ and $\ket{4}$ to the ground states $\ket{1}$ and $\ket{2}$.}
\end{figure}
In this work, we use Kerr field for precise control over the absorption of probe spectrum which leads to very tight spatial confinement of light beam. 
For this purpose, we consider a homogeneously broadened $N$-type atomic system consisting of excited states $\ket{3}$ and $\ket{4}$ and two metastable states $\ket{1}$ and $\ket{2}$,  as shown in Fig.\ref{fig:Fig1}.
The proposed model system can be experimentally realized  in $^{87}Rb$, with $\ket{1}= \ket{5S_{\frac{1}{2}},F = 2}$, $\ket{2}= \ket{5S_{\frac{1}{2}},F = 3}$, $\ket{3}= \ket{5P_{\frac{1}{2}},F^{'} = 2}$ and $\ket{4}= \ket{5P_{\frac{3}{2}},F^{'} = 4}$. 
The atomic transition $\ket{1}\leftrightarrow\ket{3}$ is coupled by a weak probe field$~(\mathcal{E}_p)$ with frequency $\omega_p$. 
A strong control field$~(\mathcal{E}_c)$ with frequency $\omega_c$, and a strong Kerr field$~(\mathcal{E}_s)$ with frequency $\omega_s$ couple the transitions $\ket{3}\leftrightarrow\ket{2}$ and $\ket{2}\leftrightarrow\ket{4}$, respectively. 
The electric fields corresponding to the three coherent fields are defined as
\begin{equation}
\label{field}
 {\vec{E}_j}(\vec{r},t)= \hat{e}_{j}\mathcal{E}_{j}(\vec{r})~e^{- i\left(\omega_j t-  k_j z\right )} + {c.c.}\,,
\end{equation}
where, $\mathcal{E}_{j}(\vec{r})$ is the slowly varying envelope, $\hat{e}_{j}$ is the unit polarization vector, $\omega_j$ is the laser field frequency and $k_j$ is the wave number of the field, respectively.  The index $j\in \{p,c,s\}$ denotes the probe, control and Kerr field, respectively.

Under the electric dipole approximation, the time-dependent Hamiltonian of the system in the presence of these three coherent fields can be written as
\begin{subequations}
\label{Hschroed}
\begin{align}
H =& H_0 + H_I\,,\\
H_0 =& \hbar\omega_{12} \ket{2}\bra{2} + \hbar\omega_{13} \ket{3}\bra{3} + \hbar\omega_{14} \ket{4}\bra{4}\,,\\
H_I =& - ( \ket{3}\bra{1} \mf{d}_{31}\cdot\mathcal{E}_pe^{- i\left(\omega_p t-  k_p z\right )}  \nonumber \\
 &  + \ket{3}\bra{2} \mf{d}_{32}\cdot\mathcal{E}_ce^{- i\left(\omega_c t-  k_c z\right )}\nonumber\\   
 &  + \ket{4}\bra{2} \mf{d}_{42}\cdot\mathcal{E}_se^{- i\left(\omega_s t-  k_s z\right )}\,+\,\text{H.c.})\,,
\end{align}
\end{subequations}
where $\mf{d}_{ij} = \bra{i}\mf{d}\ket{j}$ are dipole moments corresponding to $\ket{i}\leftrightarrow\ket{j}$ transition.
To write the Hamiltonian in the time independent form, we use the following unitary transformation
\begin{subequations}
\begin{align}
\label{unitary}
W= & e^{-\frac{i}{\hbar}Ut} \,, \\
U= &\hbar\wP\ket{3}\bra{3}+\hbar(\wP-\wC)\ket{3}\bra{3} \\ \nonumber
&+\hbar(\wP-\wC+\wM)\ket{3}\bra{3}\,.
\end{align}
\end{subequations}
Under rotating wave approximation, the effective time independent Hamiltonian takes the following form
\begin{align}
\frac{\mathcal{H}_{I}}{\hbar} & =  (\DC-\DP)\ket{2}\bra{2}+(\DC-\DP- \DM)\ket{4}\bra{4}  \notag \\
& -\DP\ket{3}\bra{3}  - \WP\ket{3}\bra{1}-\WC \ket{3}\bra{2}-\WM\ket{4}\bra{2}+\text{H.c.})\,,
 \label{Heff}
\end{align}
where the Rabi frequencies of the probe, control and the Kerr fields are defined as
\begin{subequations}
\begin{align}
\label{field}
\WP&=\frac{\vec{d}_{31}\cdot\vec{\mathcal{E}}_{\rm{p}}}{\hbar}e^{ik_p z},\WC=\frac{\vec{d}_{32}\cdot\vec{\mathcal{E}}_{\rm{c}}}{\hbar}e^{ik_c z}, \notag \\ \WM&=\frac{\vec{d}_{42}\cdot\vec{\mathcal{E}}_{\rm{s}}}{\hbar}e^{ik_s z}.\nonumber
\end{align}
\end{subequations}
The detuning of the probe, control and Kerr fields from the corresponding transition resonances are expressed as $\Delta_{p} = \omega_{p}-\omega_{31}$, $\Delta_{c} = \omega_{c}-\omega_{32}$, and $\Delta_{s} = \omega_{s}-\omega_{42}$, respectively.
We use the following Liouville equation to study the dynamics of the atomic populations and coherences, 
\begin{eqnarray}
\label{Equation7}
 \frac{\partial\rho}{\partial t} =-\frac{i}{\hbar} [\mathcal{H}_{I},\rho] + \mathcal{L}\rho\,.
\end{eqnarray}
The last term $\mathcal{L}\rho$ is Liouville operator which describes all the incoherent processes and is given by
\begin{equation}
\mathcal{L}\rho =-\sum\limits_{i=3}^4\sum\limits_{j=1}^2 \frac{\gamma_{ij}}{2}\left(\ket{i}\bra{i}\rho-2\ket{j}\bra{j}\rho_{ii}+\rho\ket{i}\bra{i}\right)\,,
\label{idecay}
\end{equation}
where $\gamma_{ij}$ corresponds to radiative decay rates from excited states $\ket{i}$ to ground states $\ket{j}$.
The equation of motion for atomic populations and coherences of the four level system can be described by the following density matrix equations:
\begin{subequations}
\label{Full_density}
\begin{align}
\dot{\rho}_{11}&= \gamma_{31}\rho_{33}+\gamma_{41}\rho_{44}+ \ii \WP^* \rho_{31}- \ii \WP \rho_{13} \,,\\
\dot{\rho}_{22}&= \gamma_{32}\rho_{33}+\gamma_{42}\rho_{44} + \ii \WC^* \rho_{32}
- \ii \WC \rho_{23}+ \ii \WM^* \rho_{42} \nonumber\\
&- \ii \WC \rho_{24} \,,\\
\dot{\rho}_{33}&= -(\gamma_{31}+\gamma_{32})\rho_{33} + \ii \WP \rho_{13}  - \ii \WP^*\rho_{31}
+ \ii \WC \rho_{23} \nonumber\\ &- \ii \WC^* \rho_{32}\,,\\
\dot{\rho}_{21}&= -\left[\gamma_{c} - \ii (\Delta_p-\Delta_c)\right]\rho_{21} - \ii \WP\rho_{23} + \ii \WC^*\rho_{31} \nonumber\\ 
&+ \ii \WM^*\rho_{41} \,,\\
\dot{\rho}_{31}&= -\left[\Gamma_{31} - \ii \Delta_p\right]\rho_{31}+ \ii \WC\rho_{21} + \ii \WP(\rho_{11} - \rho_{33})\,,\\
\dot{\rho}_{32}&= -\left[\Gamma_{32} - \ii \Delta_c\right]\rho_{32}+ \ii \WP\rho_{12} 
+ \ii \WC(\rho_{22} - \rho_{33})\nonumber\\ 
&- \ii \WM\rho_{34}\,,\\
\dot{\rho}_{34}&= -\left[\Gamma_{34} - \ii (\Delta_c-\Delta_s)\right]\rho_{34}+ \ii \WP\rho_{14} + \ii \WC\rho_{24}\nonumber\\
& - \ii \WM^*\rho_{32} \,,\\
\dot{\rho}_{41}&= -\left[\Gamma_{41} - \ii (\Delta_p-\Delta_c+\Delta_s)\right]\rho_{41}
+ \ii \WM\rho_{21}\nonumber\\ 
&- \ii \WP\rho_{43}\,,\\
\dot{\rho}_{42}&= -\left[\Gamma_{42} - \ii \Delta_s\right]\rho_{42}+ \ii \WM(\rho_{22} - \rho_{44})\nonumber\\
&- \ii \WC\rho_{43}\,,\\
\dot{\rho}_{44}&=-\dot{\rho}_{11}-\dot{\rho}_{22}-\dot{\rho}_{33}\,,\\
\dot{\rho}_{ij}&= \dot{\rho}_{ji}^*\,,
\end{align}
\end{subequations}
where, the overdots stand for time derivatives and $``^{*}"$ denotes complex conjugate. We assume that the excited states $\ket{3}$ and $\ket{4}$ decay to the ground states with equal rates, {\it i.e.,} $\gamma_{31}=\gamma_{32}=\gamma_{41}=\gamma_{42}=\gamma/2$, with $\gamma$ being the spontaneous decay rate of excited state. The coherence decay rates $\Gamma_{31}=\Gamma_{32}=\Gamma_{41}=\Gamma_{42}=\gamma/2$, and $\Gamma_{34}=\gamma$.
\subsection{Medium susceptibility}
In this section, following perturbative calculations by Dey and Agarwal \cite{Dey_07}, we derive an analytical expression for the linear susceptibility of the medium at the steady state limit. We consider the probe to be very weak so that the density matrix can be expanded to first order in probe as
  \begin{equation}
  \label{perturb}
  \rho_{_{ij}}=\rho_{_{ij}}^{(0)}+\frac{\WP}{\gamma}\rho_{_{ij}}^{(+)}+\frac{\WP^{*}}{\gamma}\rho_{_{ij}}^{(-)},
  \end{equation}
where $\rho_{_{ij}}^{(0)}$ is the zeroth order solution determined in the absence of probe field. 
The second and third term represent the first order solutions at positive and negative frequency of the probe field, respectively. 
We now substitute the above equation in Eq.~(\ref{Full_density}) and equate the coefficients of $g$ to obtain a set of 15 coupled equations. 
Next we solve these equations in the steady state limit to obtain the expression of the atomic coherence $\rho_{31}^{(+)}$. 
The linear susceptibility $\chi_{_{31}}$ of the medium at frequency $\wP$ is expressed in term of atomic coherence $\rho_{31}^{(+)}$ by the following equation
\begin{equation}
\label{Equation7}
\chi(\omega_p)= \frac{\mathcal{N}\left|d_{31}\right|^2}{\hbar\gamma}\rho_{31}^{(+)} \,,
\end{equation}
where
\begin{subequations}
\begin{align}
\label{Equation11a}\rho_{31}^{(+)}=& \frac{2\ii~\Gamma_{31}~A}{(\Gamma_{31}-\ii\DP)~A+B~\left|\WC\right|^2}\,, \\
B=& \Gamma_{41}+\ii~(\DC-\DP-\DM)\,, \\
\label{Equation11c}A=& (\gamma_{c}+\ii~(\DC-\DP))~B + \left|\WM\right|^2 \,.
\end{align}
\end{subequations}
Eq.~(\ref{Equation11a}) shows the dependence of atomic coherence on intensities of the control and the Kerr fields, respectively.
Subsequently the transverse spatial profiles of $G_{c}$ and $G_{s}$ play an important role in the manipulation of the medium susceptibility along the transverse directions.
Hence the use of an appropriate transverse profile of Kerr field creates spatial modulation of the probe absorption which enables the formation of a high contrast tuneable optical atomic waveguide.
\subsection{Beam propagation equation}
We use the Maxwell's wave equation to study the spatial evolution of the probe envelope propagating through the atomic medium which is given as 
\begin{equation}
\label{field}
\left(\vec{\nabla}^2-\frac{1}{c^2}\partial_{t}^2\right)\vec{E_{p}} = \frac{4\pi}{c^2}\partial_{t}^2 \vec{\mathcal{P}}_p.
\end{equation}
The induced polarization at $\omega_p$ is denoted by $\vec{\mathcal{P}}_{p}$ and is expressed in terms of atomic coherences as well as the susceptibility as
\begin{align}
\label{polarization}
\vec{\mathcal{P}}_p&=\mathcal{N}\left(\vec{d}_{31}\rho_{31}^{(+)}e^{-\ii\omega_p t}+c.c.\right)\notag\\
&=\left(\chi \hat{e}_{p}\mathcal{E}_pe^{-\ii\omega_p t}+c.c.\right)\,.
\end{align} 
We assume that the longitudinal variation of the probe envelope is very small as compared to the transverse variation.
In slowly varying envelope and paraxial wave approximation,  the propagation Eq.~(\ref{field}) can be transformed
to the Rabi frequency of the probe beam as 
\begin{equation}
\label{probe}
\partial_z g= \frac{\ii }{2{k_p}} \nabla^{2}_{\perp}~g + 2i{\pi}k_p{\chi}~g \,.
\end{equation}
The first terms on the right hand side of Eq.~(\ref{probe}) represents the transverse variation of the probe field and accounts for diffraction either in the medium or in free space.
The second term on the right hand side of Eq.~(\ref{probe}) lead to the dispersion and absorption of the probe beam.
To study the effect of the tranversely varying susceptibility on the probe propagation, we numerically solve Eq.~(\ref{probe}) using Fourier split-step method \cite{Bandrauk}.
However, the paraxial approximation is valid only for the light beam with $\lambda/2\pi w_0>0.1$, where $w_0$ is the beam waist \cite{Agrawal,Vaveliuk}.
Otherwise the non-paraxial term $(-i/2k_p)\partial^2_z g$ needs to be incorporated in the left hand side of Eq.~(\ref{probe}) \cite{Zhang}.
We make use of the ASRM \cite{Agrawal} to confirm the validity of paraxial approximation for propagation of narrow waist beam.
It should be borne in mind that the ASRM gives the exact solution of Maxwell's equation for a propagating beam either in free space or in a medium with spatial homogeneity.
The transverse variation of narrow waist beam at any propagation distance $z$ can be obtained by using ASRM, provided the spatial distribution of the beam at $z=0$ is known.
The spatial transverse profile of probe beam propagating in a isotropic homogeneous medium is given by
\begin{align}
\label{asrm1}
\WP(x,y,z)= \iint\limits_{-\infty}^{~~~\infty}\WP(k_{x},k_{y})e^{i[k_{x}x+k_{y}y+k_{z}z]} dk_{x} dk_{y},,
\end{align}
where
\begin{align}
\label{asrm2}
\WP(k_{x},k_{y})= \frac{1}{4\pi^{2}}\iint\limits_{-\infty}^{~~~\infty}\WP(x,y;0)e^{-i[k_{x}x+k_{y}y]} dx dy,.
\end{align}
Here $\WP(x,y;0)$ is the initial probe profile, and $k_{x}$,$k_{y}$,$k_{z}$ are the spatial frequencies of the probe. The wave vector $k_{z}$ is defined as
\begin{equation}
    k_{z}=
    \begin{cases}
      (k^{2}-(k_{x}^{2}+k_{y}^{2}))^{\frac{1}{2}}, & \text{if}\ k^{2} > (k_{x}^{2}+k_{y}^{2}) \\
      i(k^{2}-(k_{x}^{2}+k_{y}^{2}))^{\frac{1}{2}}, & \text{if}\ k^{2} < (k_{x}^{2}+k_{y}^{2})
    \end{cases}
  \end{equation}
where $k$ corresponds to wave vector in the medium and is given as
\begin{align}
\label{asrm1}
k =& k_{0}n(\omega_{p})\,, \\
n(\omega_{p})=& 1+2\pi\chi(\omega_{p})\,, \\
k_{0}=& 2\pi/\lambda_{p}.
\end{align}
\section{Results and discussion}
\begin{figure}[t]
\centering
\includegraphics[width=8.3cm, height=8.0cm]{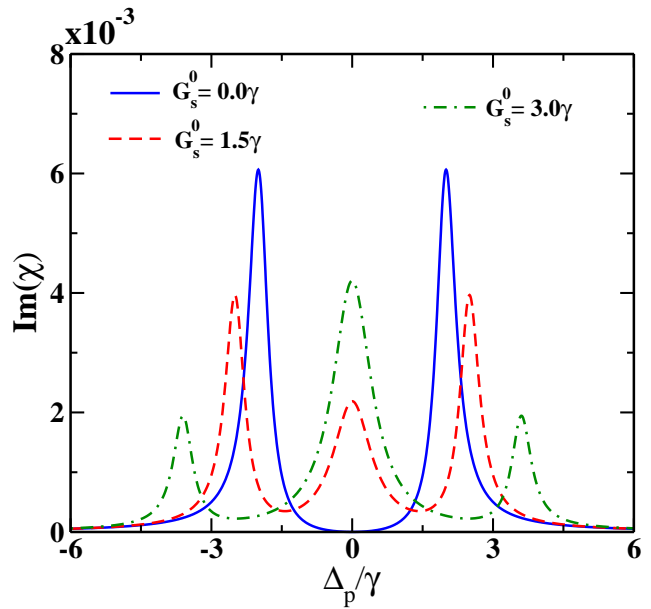}
\caption{\label{fig:Fig2} (Color online) Shows variation of imaginary part of susceptibility with probe detuning $\DP$ for different Kerr intensities. Parameters are chosen as ${\mathcal{N}}=10^{12}$ atoms/cm$^{3}$, $\Gamma_{31}=0.5{\gamma}$, $\Gamma_{41}=0.5{\gamma}$, $\DC=0.0{\gamma}$, $\DM=0.0{\gamma}$, $G_{c}^{0}=2.0\gamma$ and $\gamma_{c}=0.001{\gamma}$.}
\end{figure}
\subsection{Homogeneous field effects on linear susceptibility}
In this section, we study Eq.~(\ref{Equation7}) to demonstrate the effect of the probe absorption in the presence of both cw control and Kerr fields. 
Fig.~\ref{fig:Fig2} shows the variation of probe absorption with probe detuning at different intensities of Kerr field. 
In the absence of Kerr field, the $N$-type system reduces to a three level $\Lambda$ system wherein two arms are connected by weak probe and strong control fields, respectively. 
Hence the reduced model system exhibits a typical EIT-like feature accompanied with a transparency dip at resonance probe absorption. 
We adopt the dressed-state picture to explain the behaviour of the probe absorption \cite{Fleischhauer}. 
The two absorption peaks shown in Fig.~\ref{fig:Fig2} correspond to the transitions between bare state $\ket{1}$ to dressed states $\ket{+}=(\ket{2}+\ket{3})/\sqrt{2}$ and $\ket{-}=(\ket{2}-\ket{3})/\sqrt{2}$, respectively. 
Further, by an application of a strong Kerr field the probe transparency dip changes into an absorption peak around the resonance. 
As a result, the probe field experiences a huge absorption at the resonance.
Hence, double transparency windows are formed in place of  single transparency due to presence of peak absorption at the line center as shown in Fig.~\ref{fig:Fig2}.
Such modulation in the absorption of probe due to Kerr field is the key to confinement and guiding of light beam.   
 
Moreover, the formation of double transparency is generally due to Double-Dark resonance (DDR) effect, which can be well understood using the dressed state picture for the $N$-type system \cite{Tian}. 
In the dressed state picture, the new eigenstates of the Hamiltonian in Eq.~(\ref{Heff}) are $\ket{1}$, $\ket{0}$, $\ket{+'}$, and $\ket{-'}$.
The dressed states can be expressed in terms of the bare states $\ket{2}$, $\ket{3}$, and $\ket{4}$ as
\begin{align}
\label{dressed}
\ket{0}=& \frac{G_{c}}{\sqrt{G^{2}_c+G^{2}_s}}\ket{3}-\frac{G_{s}}{\sqrt{G^{2}_c+G^{2}_s}}\ket{4}\,, \\
\ket{+'}=& \frac{1}{\sqrt{2}}\left(\ket{2}+\frac{G_{c}}{\sqrt{G^{2}_c+G^{2}_s}}\ket{3}+\frac{G_{s}}{\sqrt{G^{2}_c+G^{2}_s}}\ket{4}\right)\,, \\
\ket{-'}=& \frac{1}{\sqrt{2}}\left(\ket{2}-\frac{G_{c}}{\sqrt{G^{2}_c+G^{2}_s}}\ket{3}+\frac{G_{s}}{\sqrt{G^{2}_c+G^{2}_s}}\ket{4}\right).
\end{align} 
The resonance absorption peak in Fig.~\ref{fig:Fig2} corresponds to the transition between $\ket{1}$ and $\ket{0}$. 
Whereas the left and right absorption peaks correspond to the transitions $\ket{1}\leftrightarrow\ket{+'}$ and $\ket{1}\leftrightarrow\ket{-'}$, respectively. 
Further, the energy separation between the dressed states $\ket{+'}$ and $\ket{-'}$ depend on both control and Kerr fields intensity and is given as $2\sqrt{G^{2}_{c}+G^{2}_{s}}$. 
Hence, with the increase in Kerr field intensity the separation between the states $\ket{+'}$ and $\ket{-'}$ increases. 
Such modulation in probe absorption due to competition between transparency and absorption, induced by control and Kerr fields, can exactly mimic the behaviour shown by photopolymers and photochromic molecules used in photolithography experiments \cite{Scott, Andrew}. 
Therefore, an $N$-level system which displays both EIT and DDR effects can have a potential application in optical lithography, image processing, etc. 
Further, the manipulation of the absorption and refractive index by the Kerr field holds the key to formation of high contrast tunable optical waveguide, which will be discussed in the next section.
\subsection{Creation of tunable waveguide}
\begin{figure}[t]
\centering
\includegraphics[width=8.4cm, height=8.0cm]{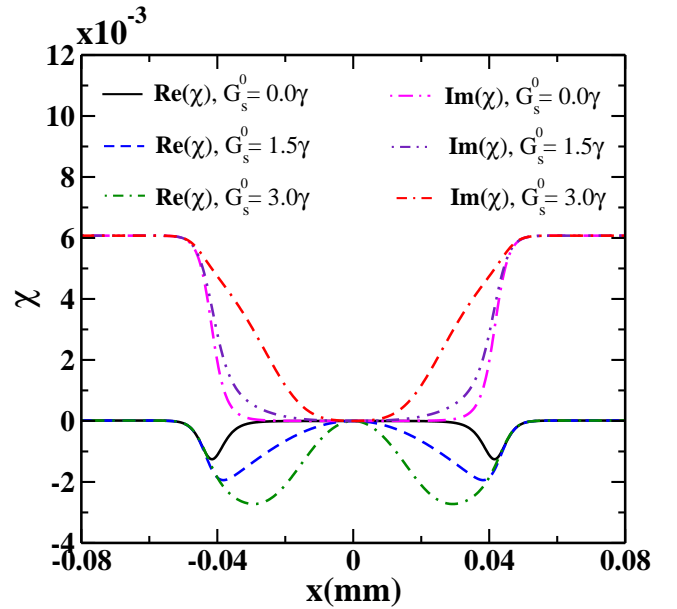}
\caption{\label{fig:Fig3} (Color online) Panel displays transverse variation of both real and imaginary parts of susceptibility at $y=0$ plane. 
Used parameters are same as in Fig.(\ref{fig:Fig2}) except the control and Kerr beam parameters are $G_{c}^{0}=2.0\gamma$, $q=1.0~$mm, $w_{c}=20~\mu$m, and $w_{s}=20~\mu$m, respectively. The probe detuning $\DP=-0.001\gamma$.}
\end{figure}

In this section, we explore the effect of spatially varying control field and Kerr field on the medium susceptibility.
In this regard, we choose control field as a Gaussian beam (LG$_{0}^{0}$) and Kerr field to be a LG beam (LG$_{m}^{l}$), respectively. 
The spatial structure of the LG beam is expressed as
\begin{subequations}
\begin{align}
&G_{j}(r,\phi,z)= G_{j}^{0}\frac{w_{j}}{w_{j}(z)}\left(\frac{r\sqrt{2}}{w_{j}(z)}\right)^{\left|l\right|}L^{l}_{m}\left(\frac{2r^{2}}{w_{j}^{2}(z)}\right)e^{ i l\phi}\nonumber \\
& e^{-\left(\frac{r^{2}}{w_{j}^{2}(z)}\right)}e^{\left(\frac{ikr^{2}}{2R_{j}(z)}\right)}e^{\left(-i(2m+l+1)\tan^{-1}\left(\frac{z}{z_{0_{j}}}\right)\right)}\,,\\
&r= \sqrt{x^{2}+y^{2}}\,, \\
&\phi=\tan^{-1}\left(\frac{y}{x}\right).
\end{align}
\end{subequations}
where $R_{j}(z)=z+(z^{2}_{0j}/z)$, and $z_{0j}=\pi w^{2}_{j}/\lambda$ is the radius of curvature and the Rayleigh length of the beams, respectively. 
The spot size of the beams is $w_{j}(z)=w_{j}\sqrt{1+((z-q)/z_{0j})^{2}}$, where, $w_{j}$ is the minimum beam waist and $q$ is the focusing point. 
The index $j\in \{c,s\}$ denotes the control and Kerr beam, respectively.
Fig.~\ref{fig:Fig3} depicts the variation of the absorption and the refractive index of the medium along the transverse direction $x$ at $y=0$ plane.
\begin{figure}[t]
\centering
\includegraphics[width=8.3cm, height=7.0cm]{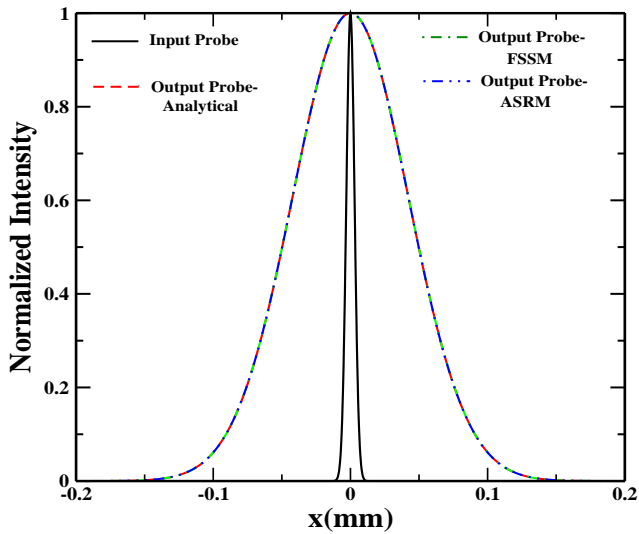}
\caption{\label{fig:Fig4} (Color online) 
Normalized intensity variation of probe envelope along transverse direction $x$ at $y=0$ plane after traversing a distance of $z=2~$mm through free space. 
The initial probe beam waist is $w_p=6\mu$m.}
\end{figure}
The results are plotted for two different cases, (i) in the presence of only control beam and (ii) in the presence of both control and Kerr beams.
In the first case, the variation of absorption and refractive index of the medium is solely dependent on the intensity  and the  spatial structure of the control beam.
The central region of  Gaussian profile of control beam permits a transparency window for the probe beam.
However, a decrease in intensity of the control beam towards the wing region gives rise to a very high probe absorption, as shown in Fig.~\ref{fig:Fig3}.
Consequently, a spatially varying probe transparency window is formed inside the atomic medium.
Interestingly, a similar behaviour of the refractive index of the medium is noticed.
It is clear from Fig.~\ref{fig:Fig3},  that the refractive index attains a maximum value at higher intensity region of the control beam which forms the core of the atomic waveguide.
Simultaneously the cladding can be manifested by gradually decreasing intensity region towards the wing of the control beam. Note that probe field is red detuned.
Figure~\ref{fig:Fig3} also shows that the induced waveguide structure consists of small refractive variation between core and cladding accompanied with a very wide core.
Such a low contrast waveguide structure fails to guide narrow waist beam due to diffraction induced distortion.
However, a suitable spatial profile of Kerr beam can significantly enhance the features of the induced waveguide.
The intensity distribution of the LG$^{1}_{0}$ Kerr beam plays an important role in controlling the absorption and refractive index of the medium.
The absorption of the medium remains unaffected at the nodal region where the intensity of the LG$^{1}_{0}$ Kerr beam vanishes.
However, the non-zero intensity of the Kerr beam contributes sharply to the increase in medium absorption towards the outer region.
As a result the transparency window gets narrow.
Further, the region with higher intensity of the Kerr beam exhibits a reduction in the refractive index between core and cladding.
Hence, a sharply varying refractive index waveguide is constituted inside the atomic medium with a narrow core structure.
We also notice from Fig.~\ref{fig:Fig3} that the width of the core can be further narrowed, and the refractive index sharply varied by using higher Kerr field intensity.
Therefore, an intense spatial Kerr profile restricts the exposure of the control profile which forms a very narrow transparency window with a high contrast refractive index. 
Thus, the spatial structre of Kerr beam plays an important role in the generation of high contrast tunable waveguide.
\begin{figure}[t]
\centering
\includegraphics[width=8.5cm, height=7.5cm]{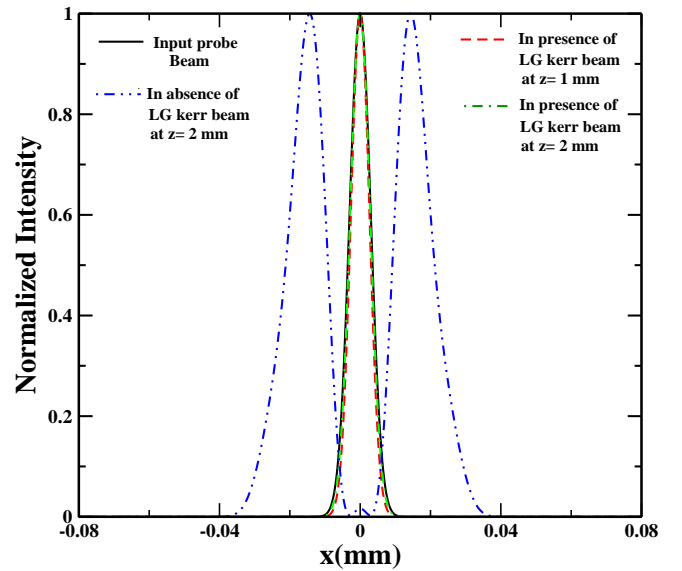}
\caption{\label{fig:Fig5} (Color online) A comparison study of narrow probe beam propagation in the absence and presence of LG$_{0}^{1}$ Kerr beams at the output of the vapor cell. 
Other parameters are same as in Fig.~\ref{fig:Fig2}.}
\end{figure}
\subsection{Beam propagation dynamics}
\begin{figure}[t]
\centering
\includegraphics[width=8.5cm, height=7.5cm]{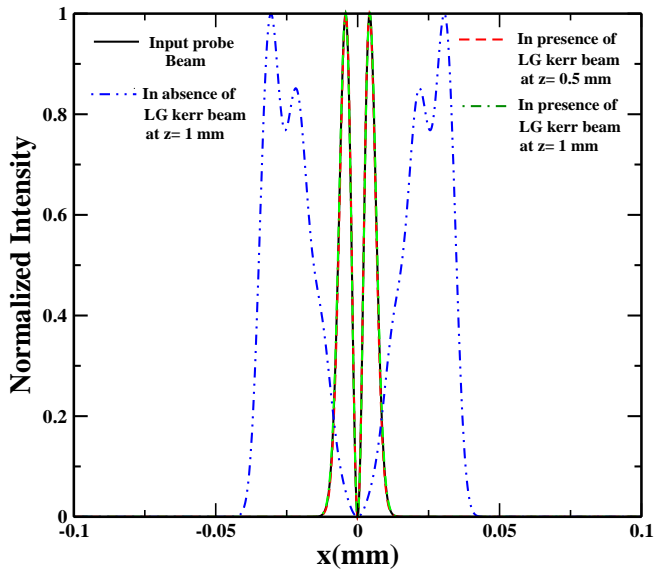}
\caption{\label{fig:Fig6} (Color online) Shows the efficient guiding of $HG_{10}$ probe in the presence of LG$^{1}_{0}$ Kerr beam. Other parameters are same as in Fig.~\ref{fig:Fig3}, except for ${\mathcal{N}}=8~\times~10^{11} atoms/cm^{3}$, $q=0.5~$mm, $w_{c}=22~\mu$m, and $w_{s}=22~\mu$m.}
\end{figure}

We now illustrate the effect of spatially varying absorption and refractive index on the probe beam propagation dynamics with different probe profiles. 
The spatial profile of the probe beam at medium entrance is given by
\begin{equation}
g(x,y,0)= g_{0}H_{n}\left(\frac{\sqrt{2}x}{w_{p}}\right)H_{m}\left(\frac{\sqrt{2}y}{w_{p}}\right)e^{-\frac{\left(x^{2}+y^{2}\right)^{2}}{w_{p}^{2}}},
\end{equation}
where $H_{n}$ and $H_{m}$ are the Hermite polynomials of order $n$ and $m$, respectively. 
Throughout our calculations, we consider probe beam width ($w_{p}$) to be 6 $\mu$m.
We consider the propagation of a Gaussian probe beam $(m=0,n=0)$. 
We numerically solve Eq.~(\ref{probe}) using Fourier split step method(FSSM) to study the propagation dynamics of probe beam having different Hermite-Gaussian modes. 
However, it is necessary to verify the validity of paraxial approximation for a beam having width of a few microns. 
For this purpose, we use ASRM as shown in Eq.~(\ref{asrm1}) to confirm the validity of paraxial approximation. 
Fig.~\ref{fig:Fig4} compares the spatial distribution of probe beams, after propagating a distance of $2$~mm in free space, which are determined by using FSSM, ASRM, and the analytical expression for paraxial Gaussian beam. 
It is clear from Fig.~\ref{fig:Fig4} that the results from FSSM matches exactly with ASRM. 
Hence, paraxial approximation holds good even for a beam with beam waist$~\sim 6 \mu m$.
It is evident that the probe beam width increases to 4$w_{p}$ after propagating a distance of $2~$mm through free space. 
This width broadening of the probe beam can be understood by considering the probe as a superposition of different plane waves. 
Each plane wave acquires different phase shift during its propagation through space. 
Hence, the resultant superposition of all the plane waves give rise to the broadening of the beam, at a given distance. 
Moreover, the broadening of the probe beam can be controlled by manipulating spatially varying refractive index of the medium. 
Such transverse variation of refractive index can be generated using a suitable control beam as shown in Fig.~\ref{fig:Fig3} \cite{Verma}.       
However, our numerical result shows that the Gaussian control beam is incapable of controlling the spreading of the narrow probe beam as seen in Fig.~\ref{fig:Fig5} and \ref{fig:Fig6}. 
The probe beam suffers severe shape distortion after propagating a distance of $2~$mm through the medium due to the presence of low contrast waveguide. 
The characteristics of the probe beam propagation through the medium drastically changes in the presence of a LG Kerr beam.  
The shape preserving probe propagation is possible due to the presence of a high contrast waveguide. 
After propagating $2~$mm $(14~Z_{r})$ through the medium, the output transmission of the probe was found to be 10$\%$. 
Further, we also show diffraction-less propagation of arbitrary modes such as $HG_{10}$ through the medium. 
Figure.~\ref{fig:Fig6} shows that $HG_{10}$ mode can propagate upto $1~$mm $(7~Z_{r})$ without any diffraction. 
The output transmission for the $HG_{10}$ mode after traversing a propagation length of $1~$mm is found to be 10$\%$. 
We also perform numerical simulations for probe beam profile with other higher order modes of HG, and found shape preserving propagation.
Hence, the spatial structure of Kerr beam plays an important role in guiding probe beams, with narrow width and arbitrary modes, to macroscopic propagation lengths.
\section{Conclusion}
We have demonstrated diffractionless propagation of narrow optical beam through tunable waveguide in a $N$-type four level system. 
We have shown how a suitably chosen spatial profile of control and Kerr beams enable us to form a high contrast atomic waveguide. 
We have also shown that contrast and tunability of the waveguide can be changed by varying the intensity of the LG Kerr beam. 
Our numerical results reveal a possibility of transferring arbitrary narrow probe image through homogeneously broadened atomic vapor to several orders of Rayleigh lengths.
\section{Acknowleadgement}
We acknowledge C. Y. Kadolkar for his helpful discussions.


\end{document}